\documentclass[prd,showpacs,twocolumn,superscriptaddress,nofootinbib,floatfix,10pt]{revtex4-2}
\usepackage{setspace}
\usepackage[utf8]{inputenc}
\usepackage{float}
\usepackage{amsmath,amssymb,amsfonts,bm}
\usepackage{graphicx}
\usepackage{multirow}
\usepackage[usenames,dvipsnames]{color}
\allowdisplaybreaks
\usepackage[
colorlinks=true,
linkcolor=blue,
breaklinks=true,
urlcolor=blue,
citecolor=blue]{hyperref}

\usepackage{orcidlink}

\usepackage{tikz}
\usetikzlibrary{arrows.meta,decorations.markings,calc}

\definecolor{green}{rgb}{0,0.6,0}

\newcommand{\be}{\begin{equation}} 
\newcommand{\ee}{\end{equation}}
\newcommand{\bea}{\begin{eqnarray}} 
\newcommand{\eea}{\end{eqnarray}}
\newcommand{\beas}{\begin{eqnarray*}} 
\newcommand{\eeas}{\end{eqnarray*}}

\newcommand{\SigD}{\Sigma_c^{(*)}\bar{D}^{(*)}}
\newcommand{\trig}{$D_{s1}^*(2860)\Lambda_c(2595)\bar{D}$}
\newcommand{\LcD}{\Lambda_{c1}\bar{D}}
\newcommand{\jp}{J/\psi p}
\newcommand{\etacn}{\eta_c p}
\newcommand{\pc}{P_c}

\newcommand{\uestc}{\affiliation{School of Physics, University of Electronic Science and Technology of China, Chengdu 611731, China}}

\newcommand{\itp}{\affiliation{Institute of Theoretical Physics, Chinese Academy of Sciences, Beijing 100190, China}}

\newcommand{\ucas}{\affiliation{School of Physical Sciences, University of Chinese Academy of Sciences, Beijing 100049, China}}

\newcommand{\scnt}{\affiliation{Southern Center for Nuclear-Science Theory (SCNT), Institute of Modern Physics,\\ Chinese Academy of Sciences, Huizhou 516000, China}}

\begin{document}
\title{Deciphering the production mechanism of the LHCb $P_c$ states}

\begin{abstract}
	The three narrow hidden-charm pentaquarks $P_c(4312)$, $P_c(4440)$, and $P_c(4457)$ observed by LHCb in $\Lambda_b^0\to J/\psi p K^-$ are widely interpreted as hadronic molecules owing to their proximity to the $\Sigma_c\bar{D}^{(*)}$ thresholds. However, heavy-quark spin symmetry predicts additional partner states that have not been seen experimentally, making the production mechanism in $\Lambda_b^0$ decays particularly important in understanding their nature. 
	In this Letter, we propose a novel production mechanism, which can naturally explain why only these three narrow LHCb $P_c$ states have been clearly observed. 
	The $P_c(4440)$ and $P_c(4457)$ are produced via the $D_{s1}^*(2860)\Lambda_c(2595)\bar{D}\pi$ box diagram leading to $K^-\Sigma_c\bar{D}^{*}$, followed by rescattering into the $K^-\jp$ final state. The box diagram develops a box singularity in the vicinity of the $P_c(4457)$, as all four states in the box can go on shell. The $P_c$ states themselves are dynamically generated through the rescattering of the $S$-wave $\Sigma_c\bar{D}^{(*)}$ to $\jp$. The $P_c(4312)$ proceeds analogously via the $D_{s0}(2590)\Lambda_c(2595)\bar{D}^*\pi$ box diagram but without a box singularity. The near-invisibility of the $\Sigma_c^*\bar{D}^{(*)}$ molecular states is attributed to the suppressed production of the $\Lambda_c(2625)$ in $\Lambda_b$ weak decays, as established by the lattice QCD calculations. This mechanism naturally reproduces the $\jp$ line shape with only two production parameters, whereas previous works required seven parameters and a finely tuned background, and provides a natural explanation for why the $\Sigma_c^*\bar{D}^{(*)}$ states are barely visible in the $\jp$ spectrum. It yields unique, falsifiable predictions, the ratio of partial widths $\Gamma(\Lambda_b^0\to P_c^+(4440/4457) K^-\to J/\psi p K^-)/\Gamma(\Lambda_b^0\to\bar D_{s1}^*(2860)\Lambda_c(2595))\approx (0.2\text{--}0.4)\%$, $\Gamma(\Lambda_b^0\to P_c^+(4312) K^-\to J/\psi p K^-)/\Gamma(\Lambda_b^0\to \bar D_{s0}(2590)\Lambda_c(2595))\approx (0.2\text{--}0.8)\%$, which can be tested at LHCb.

\end{abstract}

\author{Wen-Jia Wang\orcidlink{0000-0001-5092-0583}}
\uestc

\author{Meng-Lin Du\orcidlink{0000-0002-7504-3107}}\email{ du.ml@uestc.edu.cn}
\uestc

\author{Feng-Kun Guo\orcidlink{0000-0002-2919-2064}}\email{fkguo@itp.ac.cn}
\itp \ucas \scnt

\author{Bing Wu\orcidlink{0009-0004-8178-3015}}
\uestc

\maketitle

{\it Introduction.}--The search for multiquark exotic hadrons beyond the conventional quark model has long been a central goal of strong-interaction physics.
Clarifying the internal structure of exotic states is crucial for understanding the mechanism for quark confinement (see, e.g., the discussion in Ref.~\cite{Ji:2025hjw}).
Major progress has followed since 2003, with the discovery of the $D_{s0}^*(2317)$~\cite{BaBar:2003oey} and $X(3872)$~\cite{Belle:2003nnu}, whose properties defy conventional quark-model expectations. Among the various exotic candidates, the pentaquark states $P_c(4380)$ and $P_c(4450)$, observed by LHCb in 2015 in $\Lambda_b^0\to K^-J/\psi p$~\cite{LHCb:2015yax}, provided the first unambiguous evidence for states containing at least five quarks. An updated 2019 analysis, with an order-of-magnitude larger sample, resolved the $P_c(4450)$ into two narrow peaks, $P_c(4440)$ and $P_c(4457)$, and revealed a third, $P_c(4312)$~\cite{LHCb:2019kea}; the broad $P_c(4380)$ reported in 2015 remains to be confirmed by a full amplitude analysis.

Following the discovery of the three narrow $P_c$ states, numerous interpretations of their nature have been proposed, including hadronic molecules~\cite{Chen:2019bip,Chen:2019asm,Guo:2019fdo,Liu:2019tjn,He:2019ify,Guo:2019kdc,Shimizu:2019ptd,Xiao:2019mst,Xiao:2019aya,Wang:2019nwt,Meng:2019ilv,Wu:2019adv,Xiao:2019gjd,Voloshin:2019aut,Sakai:2019qph,Wang:2019hyc,Yamaguchi:2019seo,Liu:2019zvb,Lin:2019qiv,Wang:2019ato,Gutsche:2019mkg,Burns:2019iih,Du:2019pij,Wang:2019spc,Xu:2020gjl,Kuang:2020bnk,Peng:2020xrf,Peng:2020gwk,Xiao:2020frg,Dong:2021juy,Peng:2021hkr,Du:2021fmf,Burns:2022uiv,Pan:2022xxz,Wang:2023eng,Wu:2024bvl,Shen:2024nck,Wang:2025ecf},
compact pentaquarks~\cite{Ali:2019npk,Zhu:2019iwm,Wang:2019got,Giron:2019bcs,Cheng:2019obk,Stancu:2019qga,Kuang:2020bnk,Deng:2022vkv}, hadro-charmonia~\cite{Eides:2015dtr,Eides:2019tgv,Anwar:2018bpu}, and kinematical singularities~\cite{Guo:2015umn, Liu:2015fea,Kuang:2020bnk,Nakamura:2021qvy,Burns:2022uiv}. Among these explanations, the hadronic molecular picture is particularly attractive because the three narrow $P_c$ states lie close to the $\Sigma_c\bar{D}^{(*)}$ thresholds~\cite{Wu:2010jy}, and their pattern can be understood through the approximate heavy-quark spin symmetry (HQSS) of quantum chromodynamics (QCD). In this picture, the $P_c(4312)$ is interpreted as an $S$-wave $\Sigma_c\bar{D}$ bound state with $J^P=1/2^-$.\footnote{In Ref.~\cite{Fernandez-Ramirez:2019koa}, the $P_c(4312)$ is interpreted as a virtual state of $\Sigma_c\bar{D}$, indicating that the $\Sigma_c\bar{D}$ channel is attractive but not strong enough to form a bound state.}
The $P_c(4440)$ and $P_c(4457)$ are identified as bound states of $\Sigma_c\bar{D}^*$ with quantum numbers not uniquely fixed---they can be either $J^P=1/2^-$ and $3/2^-$, respectively, or vice versa. Importantly, HQSS predicts a total of seven bound states in the $\SigD$ systems: one $\Sigma_c\bar{D}$ bound state with $J^P=1/2^-$ identified as the $P_c(4312)$; one $\Sigma_c^*\bar{D}$ state near $4.38$ GeV with $J^P=3/2^-$; two $\Sigma_c\bar{D}^*$ bound states with $J^P=1/2^-$ and $3/2^-$, corresponding to the $P_c(4440)$ and $P_c(4457)$ (or vice versa); and three $\Sigma_c^*\bar{D}^*$ bound states with $J^P=1/2^-$, $3/2^-$, and $5/2^-$, respectively~\cite{Xiao:2013yca,Liu:2019tjn,Du:2019pij}. 
While only three of these predicted states correspond to the narrow peaks established by LHCb, there is a hint of a narrow $P_c(4380)$ as the $\Sigma_c^*\bar{D}$ molecular state with low statistical significance~\cite{Du:2019pij}. However, the three $\Sigma_c^*\bar{D}^*$ states have not been observed.
Coupled-channel amplitudes built in an HQSS-respecting effective field theory describe the $J/\psi p$ distribution well in the molecular scenario~\cite{Du:2019pij,Du:2021fmf}; including one-pion exchange (OPE) further selects $J^P=3/2^-$ and $1/2^-$ for the $P_c(4440)$ and $P_c(4457)$, respectively~\cite{Du:2019pij,Du:2021fmf}. 
The absence of signals of the $\Sigma^*_c\bar{D}^*$ states is attributed to their suppressed production rates, inferred from fits to the LHCb data of the $J/\psi p$ mass distribution, but the underlying production mechanism that could lead to such a suppression remains mysterious. The Born-Oppenheimer approach has also been applied to the $P_c$ spectrum, identifying the observed states as bound states in QCD-derived potentials~\cite{Alasiri:2025roh,Brambilla:2025xma}. In particular, the $P_c(4457)$ is assigned $J^P=5/2^-$ in Ref.~\cite{Alasiri:2025roh}, distinct from most molecular analyses.

\begin{figure}[tb]
	\begin{center}
		\includegraphics[width=1.0\linewidth]{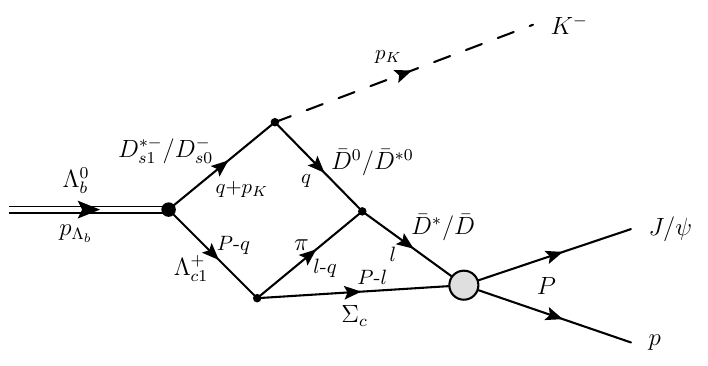}
		\caption{Proposed production mechanism of the $P_c$ states in the decay $\Lambda_b^0 \to  \jp K^-$.}
		\label{fig:FeynDiag}
	\end{center}
\end{figure}

In addition to the molecular interpretation, the $P_c(4457)$ signal may also come from a triangle singularity~\cite{LHCb:2019kea,Burns:2022uiv}. A triangle singularity arises when all intermediate particles in the triangle loop are (nearly) on their mass shell and move collinearly~\cite{Landau:1959fi, Coleman:1965xm, Bayar:2016ftu} (for a review, see Ref.~\cite{Guo:2019twa}). 
While the $P_c(4312)$ and $P_c(4440)$ structures cannot be produced by triangle singularities given realistic widths of possible intermediate states, the \trig~triangle diagram may generate a peak around the $P_c(4457)$ mass that lies close to the $\Lambda_c(2595)\bar{D}$ threshold (the $\Lambda_c(2595)$ has quantum numbers $J^P=1/2^-$ and will hereafter be denoted as $\Lambda_{c1}$ for brevity).
A possible $\LcD$ resonance interpretation with $J^P=1/2^+$ has also been explored in Refs.~\cite{Burns:2019iih,Burns:2022uiv,Wu:2024bvl}. Indeed, the $\Lambda_{c1}$ sits almost exactly at the $\Sigma_c\pi$ threshold and couples strongly to $\Sigma_c\pi$ in $S$-wave, while the $D^*$ lies very close to the $D\pi$ threshold and couples to $D\pi$ in $P$-wave. Consequently, the $S$-wave $\Sigma_c\bar{D}^*$ and $P$-wave $\LcD$ channels are connected by a one-pion exchange whose propagator is nearly on shell, which renders this transition unusually long-ranged~\cite{Geng:2017hxc,Peng:2020gwk}. This converts what na\"ively appears as a triangle diagram into a {box diagram} (see Fig.~\ref{fig:FeynDiag}), which develops a \emph{box singularity}~\cite{Landau:1959fi,Eden:1966dnq,Shen:2025qaq}. 
Such a singularity behaves as $1/\sqrt{s_0 - s}$~\cite{Gribov:2009zz}, with $s$ the Mandelstam variable and $s_0$ the singularity location, gives rise to a stronger enhancement in the vicinity of the $\Sigma_c\bar{D}^*$ threshold than the logarithmic triangle singularity, which is also present and is a subleading Landau singularity for a box diagram, and may account for the prominent signals of $P_c(4440)$ and $P_c(4457)$.

In this Letter, we propose the $D_{sJ}^{(*)}\LcD^{(*)}\pi$ box in Fig.~\ref{fig:FeynDiag} as the dominant production mechanism of the three narrow $P_c$ states observed by LHCb: $\Lambda_b^0\to \Lambda_{c1}\bar{D}_{sJ}^{(*)}\to \Sigma_c \bar{D}^{(*)} \pi K^-\to \Sigma_c\bar{D}^{(*)}K^-$, followed by the rescattering of the $S$-wave $\Sigma_c\bar{D}^{(*)}$ to the $\jp$ channels. This chain is favored because the direct production of $\Sigma_c\bar{D}^{(*)}$ in $\Lambda_b^0$ decays is color-suppressed relative to $\Lambda_b^0\to \Lambda_c^{(*)}\bar{D}_s$~\cite{Burns:2019iih,Burns:2022uiv}. 
In contrast to Refs.~\cite{Du:2019pij,Du:2021fmf}, which needed seven parameters and a finely tuned background, we will demonstrate that with merely two production parameters this mechanism can reproduce the LHCb $\jp$ line shapes and provides a natural explanation for the near-invisibility of the $\Sigma_c^*\bar{D}^{(*)}$ states, due to the suppressed production of the $\Lambda_c(2625)$ in the $\Lambda_b$ weak decays \cite{Meinel:2021rbm,Meinel:2021mdj,Du:2022rbf}.

{\it Framework.}--To describe the measured $J/\psi p$ distribution, we first construct the amplitude for the rescattering of the $S$-wave $\SigD$ into $J/\psi p$.
The transition potentials between the $\SigD$ and $J/\psi p$ systems are constrained by HQSS. The contact potentials can be derived either from heavy quark effective Lagrangians or by decomposing the spin structures of the heavy and light degrees of freedom; see, e.g., Refs.~\cite{Voloshin:2011qa,Du:2019pij}. We adopt the latter approach and expand the two-particle systems in the basis of heavy-light spin eigenstates $|s_Q\otimes j_\ell\rangle$, where $s_Q$ denotes the total spin of the heavy (anti)quark system and $j_\ell$ represents the total angular momentum of the light degrees of freedom. In this notation, the $\Sigma_c^{(*)}$ and $\bar{D}^{(*)}$ states correspond to the $\left| \frac12\otimes 1\right\rangle$ and $\left|\frac12\otimes\frac12\right\rangle$ heavy-quark spin multiplets, respectively. The $S$-wave $\SigD$ systems can then be decomposed as~\cite{Du:2019pij}
\begin{eqnarray}
	\left(\begin{array}{c}
			|\Sigma_{c}\bar{D}\rangle     \\
			|\Sigma_{c}\bar{D}^{*}\rangle \\
			|\Sigma_{c}^{*}\bar{D}^{*}\rangle
		\end{array}\right)_{\frac{1}{2}} = \left(\begin{array}{ccc}
			\frac{1}{2}          & \frac{-1}{2\sqrt{3}} & \sqrt{\frac{2}{3}} \\
			\frac{-1}{2\sqrt{3}} & \frac{5}{6}          & \frac{\sqrt{2}}{3} \\
			\sqrt{\frac{2}{3}}   & \frac{\sqrt{2}}{3}   & -\frac{1}{3}
		\end{array}\right)
	\renewcommand{\arraystretch}{1.2}
	\left(\begin{array}{c}
			|0\otimes\frac{1}{2}\rangle \\
			|1\otimes\frac{1}{2}\rangle \\
			|1\otimes\frac{3}{2}\rangle
		\end{array}\right),
	\renewcommand{\arraystretch}{1}~~\label{eq:HL1}
\end{eqnarray}
\begin{eqnarray}
	\left(\begin{array}{c}
			|\Sigma_{c}^{*}\bar{D}\rangle \\
			|\Sigma_{c}\bar{D}^{*}\rangle \\
			|\Sigma_{c}^{*}\bar{D}^{*}\rangle
		\end{array}\right)_{\frac{3}{2}} = \left(\begin{array}{ccc}
			\frac{1}{2}                   & \frac{-1}{\sqrt{3}} & \frac{1}{2}\sqrt{\frac{5}{3}} \\
			\frac{-1}{\sqrt{3}}           & \frac{1}{3}         & \frac{\sqrt{5}}{3}            \\
			\frac{1}{2}\sqrt{\frac{5}{3}} & \frac{\sqrt{5}}{3}  & \frac{1}{6}
		\end{array}\right)
	\renewcommand{\arraystretch}{1.2}
	\left(\begin{array}{c}
			|0\otimes\frac{3}{2}\rangle \\
			|1\otimes\frac{1}{2}\rangle \\
			|1\otimes\frac{3}{2}\rangle
		\end{array}\right),
	\renewcommand{\arraystretch}{1}~~\label{eq:HL2}
\end{eqnarray}
\begin{eqnarray}
	|\Sigma_{c}^{*}\bar{D}^{*}\rangle_{\frac{5}{2}}=|1\otimes\tfrac{3}{2}\rangle,\label{eq:HL3}
\end{eqnarray}
where the subscripts denote the total angular momentum $J=\frac{1}{2}$, $\frac{3}{2}$,
and $\frac{5}{2}$. For transitions between $S$-wave $\SigD$ channels, we parameterize the contact potential as
\bea
C_{j_\ell} =\delta_{s_Q s_Q^\prime}\delta_{j_\ell j_\ell^\prime}\left\langle s_Q\otimes j_\ell\left| \hat{\mathcal{H}}_I\right| s_Q^\prime\otimes j_\ell^\prime\right\rangle,
\eea
according to HQSS, with $\hat{\mathcal{H}}_I$ the effective strong-interaction Hamiltonian. The transitions between the $\SigD$ and the $J/\psi p$, as well as the $\eta_c p$ (the HQSS partner of $J/\psi p$), can be obtained analogously. We include both $S$- and $D$-waves for $J/\psi p$ and $\eta_c p$ because the center-of-mass (c.m.) momenta of the proton can be as large as $\sim 1$~GeV. In the heavy quark limit, $\left|1\otimes\frac12\right\rangle$ ($\left|0\otimes\frac12\right\rangle$) couples only to the $S$-wave $J/\psi p$ ($\eta_c p$), and $\left|1\otimes\frac32\right\rangle$ ($\left|0\otimes\frac32\right\rangle$) couples only to the $D$-wave $J/\psi p$ ($\eta_c p$)~\cite{Du:2019pij,Du:2021fmf}. Therefore, we introduce two coupling strengths~\cite{Du:2021fmf}
\bea\label{eq:inelasticcoupling}
g_S &\equiv & \left.\left.\left\langle 1\otimes \frac12 \right|\hat{\mathcal{H}}_I \right| \jp\right\rangle_S = \left.\left.\left\langle 0\otimes\frac12\right|\hat{\mathcal{H}}_I \right|\etacn \right\rangle_S,\nonumber\\
g_D k^2& \equiv & \left.\left.\left\langle 1\otimes \frac32 \right|\hat{\mathcal{H}}_I \right| \jp\right\rangle_D = \left.\left.\left\langle 0\otimes \frac32 \right|\hat{\mathcal{H}}_I \right| \etacn\right\rangle_D,
\eea
where $k$ is the magnitude of the c.m. momentum of the proton in the $\jp$ ($\etacn$) rest frame. By labeling the corresponding $\SigD$ channels on the left-hand sides of Eqs.~\eqref{eq:HL1}--\eqref{eq:HL2} with index $\alpha=1$, 2, 3, we construct the transition potentials $\mathcal{V}^{(\prime)J}_{\alpha i}$ between the $\alpha$th $\SigD$ channel and the $i$th $\jp$ ($\etacn$) channel, where $i=1,2$ denotes the $S$- and $D$-wave $\jp$ ($\etacn$), respectively. This is achieved with the help of the decompositions in Eqs.~\eqref{eq:HL1}--\eqref{eq:HL3}, as detailed in Ref.~\cite{Du:2021fmf}. Direct $\jp$ ($\etacn$) interactions are neglected because they are Okubo--Zweig--Iizuka suppressed and have been found to be weak in recent dispersive and lattice QCD studies~\cite{Skerbis:2018lew,Wu:2024xwy,Lyu:2024ttm}. As long as the direct $\jp$ ($\etacn$) interactions are neglected, the effects of the $\jp$ and $\etacn$ channels can be incorporated through their imaginary parts,
\bea
V_{\text{in},\alpha\beta}^J(E) =-\frac{ikm_p}{2\pi E}\sum_{j=1}^2\left(m_{J/\psi}\mathcal{V}_{\alpha j}^J\mathcal{V}_{\beta j}^J + m_{\eta_c}\mathcal{V}_{\alpha j}^{\prime J}\mathcal{V}_{\beta j}^{\prime J} \right), 
\eea
where $E$ is the total energy. The real parts are absorbed into a redefinition of the contact terms $C_\frac12$ and $C_\frac32$. Thus, the effective potential contains a total of four parameters and can be written as
\bea
V_{\alpha\beta}^J(E) = C_{\alpha\beta}^J+V_{\text{in},\alpha\beta}^J(E).
\eea
The contact term is expressed as 
\bea
C_{\alpha\beta}^J=\sum_{n_J} R^J_{\alpha n_J}C_{j_\ell (n_J)}(R^J)^T_{n_J\beta}\, ,
\eea
where $j_{\ell}(n_J)$ is the light-quark spin of the $n$th channel, and $R^J$ denotes the rotation matrix for a given $J$ multiplet, as defined in Eqs.~\eqref{eq:HL1}-\eqref{eq:HL3}. The unitarized scattering amplitudes among $\SigD$ can be obtained by~\cite{Oller:1998zr,Oller:2000fj}
\bea
T^J(E)=\left[1-V^J(E)\cdot G(E)\right]^{-1}\cdot V^J(E),
\eea
where $G(E)=\text{diag}\,G_\alpha(E)$ and $G_\alpha(E)$ is the two-point loop function for channel $\alpha$:
\bea\label{eq:G}
G_\alpha(E)= \int_0^{\Lambda}\frac{q^2\,\text{d}q}{2\pi^2}\frac{1}{E-m_{\text{thr},\alpha}-q^2/2\mu_\alpha+i0^+},
\eea
with $m_{\text{thr},\alpha}$ and $\mu_\alpha$ the threshold and reduced mass of channel $\alpha$, respectively. A hard cutoff $\Lambda=1$~GeV is employed to regularize the loop function. The amplitude from the $\alpha$th $\SigD$ channel with total angular momentum $J$ to $\jp$ in $S$- or $D$-wave is then given by
\bea
T_{\alpha i}^J(E) = \mathcal{V}_{\alpha i}^J + \sum_{\beta }T^J_{\alpha\beta}(E)\,G_\beta(E)\, \mathcal{V}_{\beta i}^J.
\eea

The Feynman diagram illustrating the production mechanism is depicted in Fig.~\ref{fig:FeynDiag}. The vertices for $\Lambda_b^0 \to \Lambda_{c1}\bar{D}_{s1}^*(2860)$ and $\bar{D}_{s1}^*\to \bar{D}^{(*)0} K^-$ can be described by the effective Lagrangians
\bea
\mathcal{L}_b &=& g_{\Lambda_b}\bar{\Lambda}_{c1}\, \bm{\sigma}\cdot  \bar{\bm{D}}_{s1}^{*\dagger}\,\Lambda_b^0, \\
\mathcal{L}_{s1} &=& ig_{D_{s1}^*}\,\bar{D}^{0\dagger}  \bar{\bm{D}}_{s1}^{*}\cdot \bm{\partial} K^+ + ig_{D_{s1}^*}^\prime\epsilon_{ijk}\bar{D}^{*0,i\dag}\bar{D}_{s1}^{*,j}\partial^k K^+, \nonumber
\eea
where $g_{\Lambda_b}$ and $g_{D_{s1}^*}^{(\prime)}$ are the corresponding coupling constants. The effective Lagrangian for the axial coupling of the pions to the charmed hadrons is~\cite{Wise:1992hn,Yan:1992gz,Cho:1994vg,Pirjol:1997nh}
\begin{align}
\mathcal{L}={}&\frac{g_1}{4}\langle\bm \sigma\cdot \bm u_{ab} \bar{H}_b\bar{H}_a^\dag\rangle \nonumber\\ 
& +\frac{h_2}{F_\pi}(\Sigma_c^{+\dag}\partial^0\pi^0+\Sigma_c^{0\dag}\partial^0\pi^--\Sigma_c^{++\dag}\partial^0\pi^+) \Lambda_{c1}, \nonumber
\end{align}
where $\bar{H}=-\bar{D}+\bm \sigma\cdot\bm{\bar{D}}^*$ denotes the heavy quark doublet for anticharmed mesons, $\bm u = -\nabla\Phi/F_\pi$ with $\Phi=\bm\tau\cdot\bm\pi$ and $F_\pi=92.1$ MeV. $g_1=0.57$ is determined from the $D^{*+}\to D^0\pi$ width~\cite{ParticleDataGroup:2026mpi}, and $h_2^2=0.36$ is taken from Ref.~\cite{CDF:2011zbc}.

The production of the $S$-wave $\Sigma_c\bar{D}^*$ via the box $\Lambda_{c1}\bar{D}_{s1}^*\bar{D}^0\pi$ with the intermediate $\Sigma_c\bar{D}^*$ propagators (i.e., the box and triangle parts in Fig.~\ref{fig:FeynDiag}) is given by
\begin{align}
\mathcal{P}_{\Sigma_c\bar{D}^*}={}&\frac{\sqrt{3}g_1h_2g_{D_{s1}^*}g_{\Lambda_b}m_{\Lambda_{c1}}\sqrt{2m_D}}{F_\pi^2}u_{\Sigma_c}^\dag \bm p_K\cdot\bm\sigma u_{\Lambda_b} \nonumber\\
&\times \int\frac{i^2 d^4ld^4q}{(2\pi)^8}\frac{ ({\bm l}-{\bm q})\cdot {\bm \epsilon}^*_{\bar{D}^*}E_\pi}{S_{\Sigma_c}S_{\bar{D}^*}S_\pi S_{\bar{D}}S_{\Lambda_{c1}}S_{D_{s1}^*}}\nonumber\\
={}& \mathcal{A} ~\bm p_K\cdot  {\bm \epsilon}^*_{\bar{D}^*}u_{\Sigma_c}^\dag \bm p_K\cdot\bm\sigma u_{\Lambda_b}, \label{eq:PSigDast}
\end{align}
where $S_i = p_i^2-m_i^2$ are the inverse propagators with $p_i$ and $m_i$ denoting the corresponding four-momentum and mass. Its contribution to the production amplitude of the $S$- and $D$-wave $\jp$ reads
\bea\label{eq:prod:SigDast}
P_i^J = \frac{4m_{D^*}m_{\Sigma_c} \mathcal{A}}{\sqrt{3}}p_K^2 T^J_{2i},
\eea
with $J=1/2$ and $J=3/2$.
In contrast, the production of the $S$-wave $\Sigma_c\bar{D}$ via the $\Lambda_{c1}\bar{D}_{s1}^*\bar{D}^{*}\pi$ box vanishes because
\bea
\mathcal{P}_{\Sigma_c\bar{D}} \propto g_{D_{s1}^*}^\prime u_{\Sigma_c}^\dag \bm \sigma\cdot(\bm l-\bm q)\times \bm p_K u_{\Lambda_b} \propto \bm\sigma\cdot \bm p_K\times \bm p_K =0, \nonumber
\eea
where only the relevant factors are displayed.
A nonvanishing $S$-wave $\Sigma_c\bar D$ production is obtained by replacing the $D_{s1}^*(2860)$ by the pseudoscalar $D_{s0}(2590)^-$. Its $P$-wave decay into $\bar{D}^{*0} K^-$ permits a non-zero $S$-wave projection onto $\Sigma_c\bar{D}$, thereby producing $P_c(4312)$ via the box, see e.g. in Fig.~\ref{fig:FeynDiag}, although it does not develop a box singularity since $\bar D^*\pi\to \bar D$ cannot proceed on shell. The corresponding Lagrangian is given by
\bea
\mathcal{L}_{D_{s0}}=g_{\Lambda_b}^0 \bar{\Lambda}_{c1} \bar D_{s0}^{\dagger}\,\Lambda_b^0+ ig_{D_{s0}}\bar{D}_{s0}\bar{\bm D}^{*0\dag} \cdot\bm \partial K^+,
\eea
which yields the production amplitude of the $S$- and $D$-wave $\jp$
\bea\label{eq:prod:SigD}
P^{J=1/2}_i=4m_Dm_{\Sigma_c}\mathcal{A}^\prime p_K^2 T_{1i}^{J=1/2}.
\eea
Here $\mathcal{A}^\prime$ has the same form as $\mathcal{A}$, with the masses and couplings replaced accordingly.

The two-loop function in Eq.~\eqref{eq:PSigDast} can be evaluated nonrelativistically in the $\Sigma_c\bar{D}^*$ rest frame, i.e. with $P=(E,\bm 0)$. It is given by
\begin{align}
\label{eq:loop_int}
B(E) ={}&p_K^i \int\frac{id^4q}{(2\pi)^4}\frac{1}{S_{D_{s1}^*}S_{\Lambda_{c1}}S_{\bar{D}}}\int\frac{id^4l}{(2\pi)^4}\frac{(l-q)^i E_\pi}{S_{\Sigma_c}S_{\bar{D}^*}S_\pi} \nonumber\\
\approx{}& \frac{m_\pi}{8m_{D_{s1}^*}m_{\Lambda_{c1}}m_{D}} \int_0^\Lambda \frac{q^2dq}{(2\pi)^2}\frac{I^{(1)}(q)-I(q)}{E-\omega_{\Lambda_{c1}}(\bm q)-\omega_{D}(\bm q)} \nonumber\\
  &\times   \int \frac{d \cos\theta \, \bm q\cdot\bm p_K}{E+E_K-\omega_{\Lambda_{c1}}(\bm q)-\omega_{D_{s1}^*}(\bm q+\bm p_K)},
\end{align}
where $E_K$ is the energy of $K^-$ in this frame, $\cos\theta = \bm q\cdot \bm p_K/|\bm q||\bm p_K|$, $I^{(1)}(q)$ and $I(q)$ are triangle loop functions as defined in Ref.~\cite{Guo:2010ak} (see Appendix~\ref{app:loops}), and $\omega_i(\bm p)= m_i+\bm p^2/(2m_i)$. 

The box with $\Lambda_{c1}$ in Fig.~\ref{fig:FeynDiag} feeds only $\Sigma_c\bar{D}^*$ and $\Sigma_c\bar{D}$ ($\Lambda_{c1}\to \Sigma_c^*\pi$ is $D$-wave suppressed). 
The production of $\Sigma_c^*\bar{D}^{(*)}$ can proceed via an analogous mechanism with $\Lambda_{c1}$ replaced by its heavy-quark spin partner $\Lambda_c(2625)$ with $J^P=3/2^-$. 
However, lattice-QCD calculations~\cite{Meinel:2021mdj,Meinel:2021rbm}, combined with HQSS relations~\cite{Papucci:2021pmj,Du:2022rbf}, establish that the $\Lambda_b\to\Lambda_c(2625)$ weak-decay amplitude is suppressed by a factor $\rho = 0.295(3)$ relative to $\Lambda_b\to\Lambda_{c1}$. Consequently, the box diagrams with a $\Lambda_c(2625)$ suffer a relative rate suppression of $\rho^2\approx 0.087$. More importantly, they do not exhibit any box-singularity enhancement in the energy range of interest, providing a natural explanation for the  nonobservation  of the $\Sigma_c^*\bar{D}^{(*)}$ states.

To fit the $J/\psi p$ invariant-mass distribution, we employ an incoherent smooth background to model possible contributions from misidentified events and from $\Lambda^*$ resonances coupled to $pK^-$. For this purpose, we adopt a simple parameterization:
\bea
\mathcal{B}(E) &=& b_0+b_1 E^2 + b_2 E^4,
\eea
which has three parameters fewer than that in Refs.~\cite{Du:2019pij,Du:2021fmf}.


\begin{figure}[htb]
	\centering
	\includegraphics[width=0.4\textwidth]{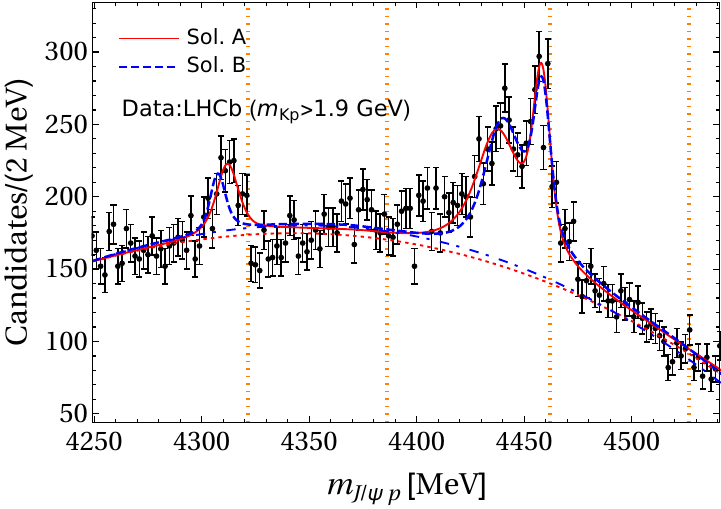}
	\caption{The fitted invariant mass distributions versus the data with $m_{Kp}>1.9$~GeV~\cite{LHCb:2019kea} for the two fit solutions. The dotted and dot-dashed curves are corresponding background. The vertical dotted orange lines denote the $\SigD$ thresholds.
A Gaussian convolution is included to account for the energy resolution of the experimental setup.	}
	\label{fig:fits}
\end{figure}

\begin{table*}[tb]
	\caption{The names of the states, their dominant channels (DCs) and thresholds, the Riemann sheet (RS) where they lie, their quantum numbers found, the pole positions, and the dimensionless couplings to their DCs, $g_\text{DC}$, from the $T$-matrix residues for Solutions $A$ and $B$. }
	\renewcommand{\arraystretch}{1.25}
	\begin{ruledtabular}
		\begin{tabular}{l |c|c| c c c | c c c}
			$\,$             & $\,$                              & $\,$       & \,          & $ \text{Solution}$ & $A$              & \,          & Solution         & $B$              \\
			\hline
			$\,$             & DC (threshold [MeV])                        & RS         & $J^{P}$     & Pole {[}MeV{]}     & $g_\text{DC}$    & $J^{P}$     & Pole {[}MeV{]}   & $g_\text{DC}$
			\tabularnewline
			\hline
			$ \pc(4312) $    & $\Sigma_{c}\bar{D}\; (4321.6)$    & RS$_{+++}$ & $\frac12^-$ & $4312(1)-5(1)i$    & $2.9(1)+0.4(2)i$ & $\frac12^-$ & $4308(4)-2(1)i$  & $3.3(2)+0.1(1)i$
			\tabularnewline
			\hline
			$P_c(4380)$ & $\Sigma_c^*\bar{D} \; (4386.2)$   & RS$_{+++}$ & $\frac32^-$ & $4375(1)-13(2)i$   & $3.1(1)+0.3(1)i$ & $\frac32^-$ & $4371(4)-12(1)i$ & $3.4(2)+0.2(1)i$
			\tabularnewline
			\hline
			$\pc(4440)$      & $\Sigma_c\bar{D}^* \; (4462.1)$   & RS$_{-++}$ & $\frac12^-$ & $4436(1)-12(3)i$   & $3.9(1)+0.5(1)i$ & $\frac32^-$ & $4438(4)-12(1)i$ & $3.9(2)+0.4(1)i$
			\tabularnewline
			\hline
			$\pc(4457)$      & $\Sigma_c\bar{D}^* \; (4462.1)$   & RS$_{-++}$ & $\frac32^-$ & $4457(2)-1(1)i$    & $2.4(1)+0.0(1)i$ & $\frac12^-$ & $4458(3)-1(1)i$  & $2.3(3)+0.0(2)i$
			\tabularnewline
			\hline
			$\pc$            & $\Sigma_c^*\bar{D}^* \; (4526.7)$ & RS$_{--+}$ & $\frac12^-$ & $4495(1)-21(4)i$   & $4.2(1)+0.6(2)i$ & $\frac12^-$ & $4524(3)-8(1)i$  & $2.0(4)+0.1(0)i$
			\tabularnewline
			\hline
			$\pc$            & $\Sigma_c^*\bar{D}^* \; (4526.7)$ & RS$_{--+}$ & $\frac32^-$ & $4513(2)-15(2)i$   & $3.2(2)+0.5(1)i$ & $\frac32^-$ & $4520(4)-13(1)i$ & $2.7(3)+0.5(1)i$
			\tabularnewline
			\hline
			$\pc$            & $\Sigma_c^*\bar{D}^* \; (4526.7)$ & RS$_{+}$   & $\frac52^-$ & $4524(1)-8(0)i$    & $1.9(3)+0.0(0)i$ & $\frac52^-$ & $4498(4)-18(1)i$ & $4.1(2)+0.4(1)i$
			\tabularnewline
		\end{tabular}\label{tab:pole}
	\end{ruledtabular}
\end{table*}

{\it Results and discussions.}--In this analysis, the widths of $\Sigma_c^{(*)}$, $D_{s1}^*(2860)$ and $D_{s0}(2590)$ are incorporated through the complex mass $m-i\Gamma/2$, with $\Gamma=1.86$ (15.76)~MeV, 159 MeV, and 89 MeV, respectively. A Gaussian convolution is included to account for the experimental energy resolution~\cite{LHCb:2019kea}. We fit the $m_{\jp}$ distribution data with the cut $m_{Kp}>1.9$~GeV~\cite{LHCb:2019kea}, which is applied to effectively remove the crossed-channel $\Lambda^*$ contributions.

Similar to the contact-potential results of Refs.~\cite{Du:2019pij,Du:2021fmf}, which yielded two nearly degenerate solutions that adequately describe the data, our new mechanism also yields two distinct solutions, denoted as A and B following Refs.~\cite{Liu:2019tjn,Du:2019pij,Du:2021fmf}, with $\chi^2/\text{d.o.f}=1.37$ and 1.57, respectively (the parameter values are given in Appendix~\ref{app:results}). The corresponding results are displayed in Fig.~\ref{fig:fits}. 
Even with a simple background parameterization, our mechanism reproduces the experimental data satisfactorily. Notably, the descriptions of the $P_c(4440)$ and $P_c(4457)$ peaks are significantly improved compared to previous works~\cite{Du:2019pij,Du:2021fmf}, which employed seven free parameters for the weak production $\Lambda_b\to K^-\SigD$ and a background parameterization with an additional broad Breit-Wigner function containing three more parameters. The successful agreement with the experimental data supports both the molecular interpretation of the $P_c$ states as $\SigD$ hadronic molecules and the proposed $\Lambda_{c1}\bar{D}_{s1(0)}^{(*)}\bar{D}^{(*)}\pi$ box-diagram production mechanism.

With the fitted amplitude, we can search for poles. In a coupled-channel system with $n$ distinct thresholds, there are $2^n$ Riemann sheets, labeled as $\text{RS}_{\pm\pm\dots\pm}$, where each subscript denotes the sign of the imaginary part of the c.m. momentum for the $\alpha$th channel with the channels ordered by increasing thresholds. Consequently, for $J=1/2$ and $J=3/2$ there are $8$ RSs, while for $J=5/2$ only $2$ sheets exist. In an $n$-channel case, only $n+1$ of the $2^n$ RSs are of primary interest, as they are directly connected to the physical region by crossing the unitarity cut. Poles located on these sheets have a significant impact on physical observables. 
When searching for poles corresponding to the $P_c$ states, the $\jp$ and $\eta_c p$ channels are taken to be on their unphysical sheets. Table~\ref{tab:pole} summarizes the pole positions, their associated sheets, and the dimensionless effective couplings (derived from residues of the $T$ matrix via $g_\alpha g_\beta = \lim_{E\to E_\text{pole}}(E^2-E_\text{pole}^2)T^J_{\alpha\beta}(E)$) to the dominant channels (DCs), defined as those with the largest couplings. These results are consistent with those reported in Ref.~\cite{Du:2021fmf}. All of the $\SigD$ states are found below their respective thresholds and are thus identified as quasi-bound states.

The $P_c(4312)$ couples dominantly to $\Sigma_c\bar{D}$ with $J^P=\frac12^-$ and can be interpreted as a $\Sigma_c\bar{D}$ bound state. The poles corresponding to the $P_c(4440)$ and $P_c(4457)$ reside on $\text{RS}_{-++}$ and are identified as $\Sigma_c\bar{D}^*$ bound states with $J^P=\frac12^-$ and $\frac32^-$, respectively, in Solution A; for Solution B, the assignments are reversed ($J^P=\frac12^-$ for $P_c(4457)$ and $J^P=\frac32^-$ for $P_c(4440)$). The observed signals of these two structures arise from the combined effects of the poles and the cusp generated by the box-triangle two-loop function (Eq.~\eqref{eq:loop_int}) at the $\Sigma_c\bar{D}^*$ threshold. As in Refs.~\cite{Du:2019pij,Du:2021fmf}, a narrow $P_c(4380)$ state with $J^P=\frac32^-$ emerges as a $\Sigma_c^*\bar{D}$ bound state. Its existence is a natural consequence of HQSS and is hinted at by the experimental data \cite{Du:2019pij,Du:2021fmf}. 
Its significance is much weaker than that of the three narrow $P_c$ states, there is no visible stucture here owing to the suppressed $\Lambda_{c}(2625)$ production. 
Similarly, three $\Sigma_c^*\bar{D}^*$ poles are found on RS$_{--+}$ for $J=1/2$, 3/2, and on RS$_+$ for $J= 5/2$. The mass ordering of these three $\Sigma_c^*\bar{D}^*$ states follows a pattern analogous to that of the $\Sigma_c\bar{D}^*$ states: for Solution A, $m_{1/2^-}<m_{3/2^-}<m_{5/2^-}$ , while for Solution B the ordering is reversed. It has been demonstrated in Ref.\cite{Du:2021fmf} that, when the OPE is included in the $\SigD$ interactions, Solution A acquires a strong cutoff dependence and is thus disfavored, while Solution B exhibits a much milder cutoff dependence and is thus considered more natural. In the present Letter we focus solely on the production mechanism of the $P_c$ states. We note that Ref.\cite{Du:2021fmf} assumes seven independent pointlike production vertices, and the cutoff dependence arises at the level of the $\SigD\to\jp$ rescattering amplitude rather than from the production mechanism. Since our new production mechanism feeds into this same rescattering amplitude, the same cutoff-dependence behavior is expected to persist here.

We have also attempted to explicitly include the $P$-wave $\Lambda_{c1}\bar{D}$ channel via contact potentials, treated as free parameters in the fit to the data. The results indicate that the $P$-wave $\LcD$ channel plays only a minor role for the $P_c(4440)$ and $P_c(4457)$ states, consistent with the findings of Ref.~\cite{Yalikun:2021bfm}.

In the production mechanism illustrated in Fig.~\ref{fig:FeynDiag}, the partial widths for $\Lambda_b^0\to P_c^+(4440/4457) K^-\to \jp K^-$ and $\Lambda_b^0\to P_c^+(4312) K^-\to \jp K^-$ are governed dominantly by Eq.~\eqref{eq:prod:SigDast} and~\eqref{eq:prod:SigD}, respectively. Our numerical checks indicate that the coupled-channel contributions from Eq.~\eqref{eq:prod:SigDast} to the $P_c(4312)$ production, and from Eq.~\eqref{eq:prod:SigD} to the $P_c(4440/4457)$ production, are negligible. Although the parameters $g_{\Lambda_b} g_{D_{s1}^*}$ and  $g_{\Lambda_b}^0  g_{D_{s0}}$ cannot be determined solely from fits to the $\jp$ mass distributions due to an unknown overall normalization, the coupling $g_{D_{s1}^*}$ can be inferred from the decay $D_{s1}^*(2860)\to DK$, whose partial width is estimated to be around $80$~MeV~\cite{Guo:2011dd}. This allows us to predict the ratio
\begin{equation}\label{eq:R}
	\frac{\Gamma(\Lambda_b^0\to P_c^+(4440/4457) K^-\to \jp K^-)}{\Gamma(\Lambda_b^0\to \bar D_{s1}^*(2860)\Lambda_{c1}^+)} \approx (0.2\text{--}0.4)\%.
\end{equation}
Similarly, by assuming the partial width of $D_{s0}(2590)\to D^*K$ to be half of its total width, one has
\begin{equation}\label{eq:R2}
	\frac{\Gamma(\Lambda_b^0\to P_c^+(4312) K^-\to \jp K^-)}{\Gamma(\Lambda_b^0\to \bar D_{s0}(2590)^-\Lambda_{c1}^+)} \approx (0.2\text{--}0.8)\%.
\end{equation}
These unique predictions are direct consequences of the proposed production mechanism and thus falsifiable with measurements at LHCb.


{\it Summary and outlook.}--In summary, we have proposed that the three narrow LHCb $P_c$ states are produced via the $\Lambda_{c1}D_{s1(0)}^{(*)}\bar{D}^{(*)}\pi$ box diagrams, followed by the rescattering of the $S$-wave $\Sigma_c\bar{D}^{(*)}$ into the $\jp$ channels. Within a unitary coupled-channel framework constrained by HQSS, this mechanism reproduces the LHCb $J/\psi p$ line shape with only two production parameters---in contrast to the seven employed in~\cite{Du:2019pij,Du:2021fmf}---and three fewer background parameters. 
The mechanism has a box-singularity enhancement near the $\Sigma_c\bar{D}^*$ threshold, naturally explaining why the $P_c(4440)$ and $P_c(4457)$ signals are pronounced. It reinforces the molecular interpretation of the $P_c$ states. 
Crucially, the mechanism also explains the nonobservation of the $\Sigma_c^*\bar{D}^{(*)}$ states, since the $\Lambda_c(2625)$ production from $\Lambda_b$ weak decays that feeds them is heavily suppressed. 
It yields sharp, falsifiable predictions, $\Gamma(\Lambda_b^0\to P_c^+(4440/4457) K^-\to J/\psi p K^-)/\Gamma(\Lambda_b^0\to \bar D_{s1}^*(2860)\Lambda_c(2595))\approx (0.2\text{--}0.4)\%$, and $\Gamma(\Lambda_b^0\to P_c^+(4312) K^-\to J/\psi p K^-)/\Gamma(\Lambda_b^0\to \bar D_{s0}(2590)\Lambda_c(2595))\approx (0.2\text{--}0.8)\%$, which can be tested at LHCb. More broadly, our results highlight that the production dynamics can play a decisive role in establishing the nature of near-threshold exotic hadrons.

\medskip

\acknowledgments
This work is supported in part by the National Natural Science Foundation of China (NSFC) under Grants No.~12675095, No.~12547111, No. 12125507,  and No. 12447101; by the National Key R\&D Program of China under Grant No. 2023YFA1606703; and by the Chinese Academy of Sciences (CAS) under Grant No.~YSBR-101. MLD gratefully acknowledges the support of the Peng Huan-Wu Visiting Professorship and the hospitality of the Institute of Theoretical Physics, CAS, where part of this work was done.

\bibliography{Pc_refs.bib}

\begin{onecolumngrid}
	\begin{appendix}

	\section{Loop functions}
	\label{app:loops}
		The nonrelativistic two-point loop function reads
		\bea
		G(E) &=& \int_0^{q_\text{max}}\frac{q^2\,\text{d}q}{2\pi^2}\frac{1}{E-m_\text{thr}-q^2/2\mu+i0^+ }=-\frac{\mu }{\pi^2}\left[ q_\text{max}+\frac{p}{2}\left(i\pi-\log\frac{q_\text{max}+p}{q_\text{max}-p}\right)\right],
		\eea
		where $p=\sqrt{2\mu(E-m_\text{thr})}$. The basic three-point scalar loop function is defined as
		\bea
		I(q)\equiv i\int \frac{\text{d}^4l}{(2\pi)^4}\frac{1}{(l^2-m_1^2)\left[ (P-l)^2-m_2^2\right]\left[ (l-q)^2-m_3^2\right]},
		\eea
		where $m_1$, $m_2$, $m_3$ correspond to the masses of $\bar{D}^{(*)}$, $\Sigma_c$, and $\pi$, respectively,  in Fig.~\ref{fig:FeynDiag}; $P$ is the four-momentum of the $S$-wave $\Sigma_c\bar{D}^{(*)}$ system. Nonrelativistically, in the rest frame of the $\Sigma_c\bar{D}^{(*)}$, $P^\mu=\left(E,\bm{0}\right)$, the integral can be worked out by performing the contour integration over $l^0$ and then integrating over the spatial components. The result reads~\cite{Guo:2010ak}
		\bea
		I(q)=\frac{\mu_{12}\mu_{23}}{16\pi m_1m_2m_3}\frac{1}{\sqrt{a}}\left[\tan^{-1}\frac{c^\prime-c}{2\sqrt{ac}}+\tan^{-1}\frac{2a+c-c^\prime}{2\sqrt{a(c^\prime-a)}}\right],
		\eea
		where $\mu_{ij}=m_i m_j/(m_i+m_j)$,
		\bea
		a=\left(\frac{\mu_{23}}{m_3}\right)^2\vec{q\,}^2,\quad c=2\mu_{12}b_{12}, \quad c^\prime=2\mu_{23}b_{23}+\frac{\mu_{23}}{m_3}\vec{q\,}^2,
		\eea
		with $b_{12}=m_1+m_2-E$ and $b_{23}=m_2+m_3+E_q-E$, where $E_{q}$ is the energy of $\bar{D}^{(*)}$ with three-momentum $\vec{q}$, i.e., $E_q=m_{D^{(*)}}+{\vec{q\,}}^2/2m_{D^{(*)}}$. The vector three-point loop integral is defined as
		\bea
		q^iI^{(1)}(q) \equiv i\int \frac{\text{d}^4l}{(2\pi)^4}\frac{l^i}{(l^2-m_1^2)\left[ (P-l)^2-m_2^2\right]\left[ (l-q)^2-m_3^2\right]}.
		\eea
		The expression of $I^{(1)}(q)$ is given by
		\bea
		I^{(1)}(q)=\frac{\mu_{23}}{m_3 a}\left[\frac{\mu_{12}\mu_{23}}{16\pi m_1m_2m_3}(\sqrt{c}-\sqrt{c^\prime-a})+\frac12(c^\prime-c)I(q)\right].
		\eea

\begin{figure}[tbh]
			\centering
			\includegraphics[width=0.5\textwidth]{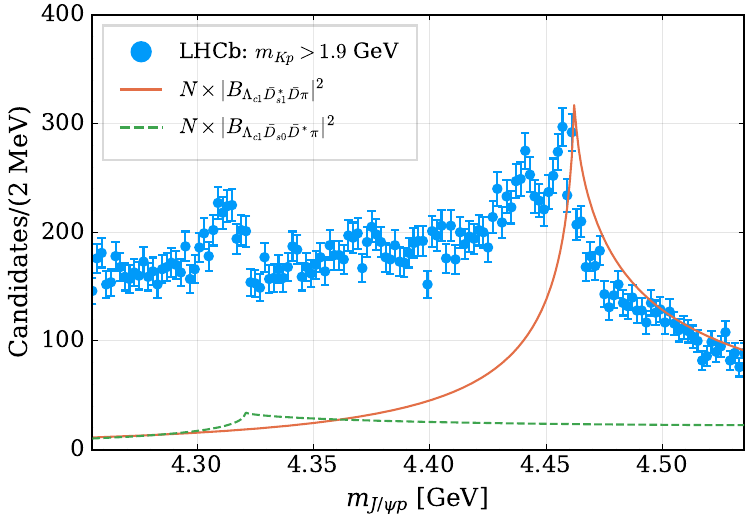}
			\caption{Comparison between the LHCb $\jp$ invariant-mass distribution data with $m_{Kp}>1.9$~GeV~\cite{LHCb:2019kea} and the box-triangle two-loop functions of $B_{\Lambda_{c1}\bar{D}_{s1}^*(2860)\bar{D}\pi}$ and $B_{\Lambda_{c1}\bar{D}_{s0}(2590)\bar{D}^*\pi}$. A normalization factor $N$ has been introduced.}
			\label{fig:Box:data}
\end{figure}
Figure~\ref{fig:Box:data} compares the box-triangle two-loop function $B(E)$ in Eq.\eqref{eq:loop_int} with the LHCb measurement of the $\jp$ mass distribution under the cut $m_{Kp}>1.9$ GeV \cite{LHCb:2019kea}, where a normalization factor $N$ is introduced for comparison. The box-triangle loop function for the $S$-wave $\Sigma_c\bar{D}$, i.e., $B_{\Lambda_{c1}\bar{D}_{s0}(2590)\bar{D}^*\pi}$, does not generate a box singularity and thus only exhibits a threshold cusp, whereas that for the $\Sigma_c\bar{D}^*$ develops a box singularity, leading to a strong enhancement near the $\Sigma_c\bar{D}^*$ threshold.

\section{Results of the fits}
\label{app:results}

This study was performed neglecting isospin symmetry breaking effects. The masses of particles used in the calculation are taken as
\begin{center}
\begin{alignat*}{4}
m_{\Lambda_b^0} &= 5.6196\ \text{GeV}, \quad &
m_K             &= 0.4956\ \text{GeV}, \quad &
m_{J/\psi}      &= 3.0969\ \text{GeV}, \quad &
m_{\Sigma_c}    &= 2.4535\ \text{GeV}, \\
m_p             &= 0.9383\ \text{GeV}, \quad &
m_{\eta_c}      &= 2.9839\ \text{GeV}, \quad &
m_\pi           &= 0.1380\ \text{GeV}, \quad &
m_{\Sigma_c^*}  &= 2.5181\ \text{GeV}, \\
m_D             &= 1.8680\ \text{GeV}, \quad &
m_{D^*}         &= 2.0086\ \text{GeV}, \quad &
m_{\Lambda_{c1}}&= 2.59225\ \text{GeV}, \quad &
m_{D_{s1}^*}    &= 2.859\ \text{GeV}, \\
m_{D_{s0}}      &= 2.591\ \text{GeV}.
\end{alignat*}
\end{center}

The best-fit parameters for the results shown in Fig.~\ref{fig:fits} are collected in Table~\ref{tab:parameters}.

\begin{table*}[tbh]
	\caption{Best-fit parameters for Solutions A and B. Here $k_0$ is the proton momentum in the $\jp$ c.m. frame at the reference c.m. energy $(m_{D}+m_{D^*}+m_{\Sigma_c}+m_{\Sigma_c^*})/2$. }\label{tab:parameters}
\begin{tabular}{l|c|c| c | c| c}
\hline
\hline
Solution & $C_{1/2}$ [GeV$^{-2}$] & $C_{3/2}$ [GeV$^{-2}$] & $g_S$ [GeV$^{-2}$] & $g_D k_0^2$ [GeV$^{-2}$] & $g_{\Lambda_b^0}g_{D_{s0}}/g_{\Lambda_b}g_{D_{s1}^*}$ \\
\hline
A & $-14.33(18)$ & $-9.92(14)$ & 3.75(25) & 0.00(63) & 1.92(12) \\
\hline
B & $-9.70(18)$& $-13.26(13)$ & 0.00(25) & 2.71(27) &  $-0.60(12)$ \\
\hline
\hline
\end{tabular}
\end{table*}

	\end{appendix}

\end{onecolumngrid}

\end{document}